\documentclass[prd,aps,preprint,draft,showpacs]{revtex4}
\usepackage{epsf}
\usepackage[dvips]{color}
\usepackage{dcolumn}% Align table columns on decimal point
\begin{document}
% declarations for front matter
\title{Topological susceptibility with the improved Asqtad action}
\author{Claude Bernard}
\affiliation{
Department of Physics, Washington University, St.~Louis, MO 63130, USA
}
\author{
Thomas DeGrand and Anna Hasenfratz
}
\affiliation{
Physics Department, University of Colorado, Boulder, CO 80309, USA
}
\author{
Carleton DeTar and James Osborn
}
\affiliation{
Physics Department, University of Utah, Salt Lake City, UT 84112, USA
}
\author{
Steven Gottlieb
}
\affiliation{
Department of Physics, Indiana University, Bloomington, IN 47405, USA
}
\author{
Eric Gregory and Doug Toussaint
}
\affiliation{
Department of Physics, University of Arizona, Tucson, AZ 85721, USA
}
\author{
Alistair Hart
}
\affiliation{
School of Physics, University of Edinburgh, King's Buildings, Edinburgh, EH9 3JZ, UK
}
\author{
Urs M.~Heller
}
\affiliation{
American Physical Society, One Research Road, Box 9000, Ridge, NY 11961, USA
}
\author{
James Hetrick
}
\affiliation{
Physics Department, University of the Pacific, Stockton, CA 95211, USA
}
\author{
Robert L.~Sugar
}
\affiliation{
Department of Physics, University of California, Santa Barbara, CA 93106, USA
}
\date{\today}

\begin{abstract}
Chiral perturbation theory predicts that in quantum chromodymamics
light dynamical quarks suppress the topological (instanton)
susceptibility.  We investigate this suppression through direct
numerical simulation using the Asqtad improved lattice fermion action.
This action holds promise for carrying out nonperturbative simulations
over a range of quark masses for which chiral perturbation theory is
expected to converge.  To test the effectiveness of the action in
capturing instanton physics, we measure the topological susceptibility
as a function of quark masses with $2 + 1$ dynamical flavors.  Our
results, when extrapolated to zero lattice spacing, are consistent
with predictions of leading order chiral perturbation theory.

Included in our study is a comparison of three methods for analyzing
the topological susceptibility: (1) the Boulder hypercubic blocking
technique with the Boulder topological charge operator, (2) the more
traditional Wilson cooling method with the twisted plaquette
topological charge operator and (3) the improved cooling method of de
Forcrand, Perez, and Stamatescu and their improved topological charge
operator.  We show in one comparison at nonzero lattice spacing that
the largest difference between methods (1) and (2) can be attributed
to the operator, rather than the smoothing method.
\end{abstract}

\maketitle

\section{Introduction}

Chiral perturbation theory predicts the behavior of the topological
susceptibility in the limit of small quark mass.  For improved fermion
actions such as Asqtad\cite{ref:asqtad} that lack complete chiral
symmetry at nonzero lattice spacing, reproducing this prediction is a
particularly challenging test.

The Asqtad improvement adds local three-, five-, and seven-link terms to
the standard staggered fermion action to eliminate tree-level lattice
artifacts to order $a^2$ \cite{ref:asqtad}.  Here we use it in
conjunction with the one-loop improved Symanzik gauge action.  This
action has proven to be highly successful in determining the masses of
the light hadrons \cite{ref:asqtadspectrum} and a variety of
quarkonium masses and meson decay parameters \cite{ref:precision}.  It
appears that the most satisfactory agreement with experimental values
is achieved so far for those quantities that are well behaved in the
chiral limit.

Whether improvement is successful in reducing lattice artifacts
clearly depends on the observable.  The Asqtad quark-gluon vertex is
not as smooth as that of the more elaborate HYP action \cite{ref:HYP}
and zero modes are not treated as rigorously as with the more
expensive domain wall and overlap actions \cite{ref:domain,
ref:overlap}.  Through a rougher vertex, quark propagation might be
influenced by small instanton-like dislocations.  With imprecise zero
modes, at small quark mass the fermion determinant may fail to
suppress adequately configurations with nonzero topological charge.

The gluonic measurement of topological charge by summing the charge
density is sensitive to the choice of both the discretization of the
charge density operator and the smoothing or cooling method.
Consequently, we found it instructive to compare three methods for
measuring the charge:

\begin{enumerate}
  \item the Boulder definition of the topological
charge density \cite{ref:DHK.SU2} with smoothing through hypercubic
blocking \cite{ref:HYPblock}
  \item the more traditional combination of
measuring the charge density through the twisted plaquette operator
and cooling by minimizing the Wilson action
\cite{ref:twplq,ref:Wilsoncool}, and
  \item measuring the charge density with the five-loop improved
operator and cooling with a five-loop improved gauge action
\cite{ref:fiveLi}.
\end{enumerate}
Throughout, we shall abbreviate these methods with ``Boulder/HYP'',
``TwPlaq/Wilson'', and ``5Li/5Li'', respectively.

No previous study of the topological susceptibility has shown
satisfactory agreement with the predictions of chiral perturbation
theory at quark masses much smaller than the strange quark mass.
Until recently \cite{ref:CPPACS2001,ref:Anna2001,ref:HT2001}, even the
expected suppression of the susceptibility at small dynamical quark
mass has been difficult to detect
\cite{ref:previous,ref:kovacs2001}. We argue that a combination of
improvements in the lattice action, the smoothing (cooling) technique,
the topological charge operator, and an $O(a^2)$ extrapolation to the
continuum lead to plausible agreement with lowest order chiral
perturbation theory for small quark masses.

This article is organized as follows.  In Sec.~\ref{sec:toposusc} we
introduce notation and review the predictions of chiral perturbation
theory.  Three methods for measuring the topological charge are
compared and discussed in Sec.~\ref{sec:method}.  Results are
presented in Sec.~\ref{sec:results} and discussed in
Sec.~\ref{sec:discussion} and conclusions are given in
Sec.~\ref{sec:conclusions}.

Preliminary results of this study were reported at the Lattice 2002
conference \cite{ref:suscept02}.

\section{Topological Susceptibility}
\label{sec:toposusc}

The topological charge $Q$ is the integral of the topological charge
density $\rho$, which is in turn defined in terms of the Euclidean
color gauge field $F^a_{\mu\nu}$ and its dual $\tilde F^a_{\mu\nu}$,
through
\begin{equation}
  Q =  \int\rho\, d^4x = \frac{1}{32 \pi^2}\int F^a_{\mu\nu} \tilde F^a_{\mu\nu} d^4x.
\label{eq:Q}
\end{equation}
On a lattice of Euclidean space-time volume $V$ the topological
susceptibility is the mean fluctuation in the topological charge,
\begin{equation}
  \chi = \frac{\langle Q^2 \rangle}{V}
\label{eq:suscept}
\end{equation}
Chiral perturbation theory predicts \cite{ref:topochiral,ref:LS1992}
that at large $\langle Q^2 \rangle$ (i.e.\ large values of $mV\Sigma$,
the product of the lightest quark mass, the lattice volume, and the
chiral condensate parameter), and at sufficiently small quark masses,
the topological susceptibility is related to the quark masses through
\begin{equation}
  \chi = \frac{\Sigma}{\left( 1/m_u + 1/m_d + 1/m_s\right)}.
\end{equation}
For three flavors with $m_u = m_d = m_{u,d}$ we may use the PCAC
relation between the up and down quark mass, pion decay constant, and
the chiral condensate to write the susceptibility as
\begin{equation}
 \chi = \frac{f_\pi^2 m_\pi^2}{4(1 + m_{u,d}/2 m_s)},
\label{eq:chiral}
\end{equation}
showing that it vanishes linearly in the square of the pion mass in
the chiral limit.

At infinite quark mass (lattice quenched approximation) the
susceptibility is finite, requiring negative curvature corrections
in Eq (\ref{eq:chiral}), leading asymptotically to a constant
\cite{ref:LS1992,ref:HT2001,ref:Durr}.  Chiral perturbation theory,
however, is not expected to converge for masses greater than $m_s$, so
it provides no guidance there.

We test Eq (\ref{eq:chiral}) against our measured susceptibility, pion
mass \cite{ref:spectlat02}, and pion decay constant
\cite{ref:precision} over a range of light quark masses.

\section{Measuring the Topological Charge on the Lattice}
\label{sec:method}

In this section we compare three methods for measuring the topological
charge.  All three methods first smooth out ultraviolet fluctuations
and then measure the topological charge density with a local
discretized operator.  All are equivalent in the continuum limit.
However, some perform better at nonzero lattice spacing.  To
understand the comparison it is first useful to briefly review the
effects of discretization on the topological susceptibility.

\subsection{Discretization effects}
\label{subsec:artifacts}

The topological susceptibility measured on the lattice differs from
the continuum value because of discretization errors in the
measurement process and in the lattice action itself.  The principal
errors introduced in the measurement process are these:

\begin{enumerate}

\item
Instantons with a core size of the order of or less than the lattice
spacing are excluded altogether,

\item
Small instantons shrink and are erased by prolonged smoothing,

\item
Intermediate sized instantons have a topological charge less than
unity, owing to the discretization of the charge density operator,

\item
There are `dislocations' on the lattice: ultraviolet fluctuations that
masquerade as topological charges,

\end{enumerate}

The principal errors introduced by the action are these:

\begin{enumerate}
\item
For unquenched gauge configurations the staggered Dirac matrix has
inexact zero modes so fails to fully ``see'' and suppress topological
fluctuations adequately, and

\item
Lattice artifact flavor (taste) symmetry breaking fails to account
for light quark flavors correctly.

\end{enumerate}

\subsection{Comparison of operators and smoothing methods}

We consider the following topological charge density operators:

\begin{enumerate}
  \item TwPlaq: The original twisted plaquette operator, defined as an
        eight-link path with a displacement sequence
        $\{x,y,-x,-y,z,t,-z,-t\}$ plus rotations.
  \item 5Li: The five-loop improved operator of de Forcrand, Perez, and
        Stamatescu \cite{ref:fiveLi}, built from a linear combination
        of five operators in the form of the twisted plaquette, but
        with the plaquettes replaced by various $m\times n$
        rectangular Wilson loops.
  \item Boulder: The lattice approximation developed for SU(2) by
        DeGrand, Hasenfratz, and Kovacs \cite{ref:DHK.SU2} and refined
        for SU(3) by Hasenfratz and Nieter \cite{ref:HN.SU3}.  It
        involves a combination of two contorted Wilson loop operators
        in the fundamental and adjoint representations of SU(3), both
        defined on closed ten-link paths described by unit lattice
        vector displacements in the sequence
        $\{x,y,z,-y,-x,t,x,-t,-x,-z\}$ and
        $\{x,y,z,-x,t,-z,x,-t,-x,-y\}$ plus rotations and cyclic
        permutations.  This operator was optimized to reduce lattice
        corrections for small instantons with radii close to the
        lattice spacing $R \approx a$.
\end{enumerate}
All three operators are equivalent in the continuum limit, but they
are subject to different discretization effects.  Reference
\cite{ref:HetrickLat02} makes a comparison of methods 2 and 3.

\begin{figure}[ht]
 \epsfxsize=100mm
 \epsfbox{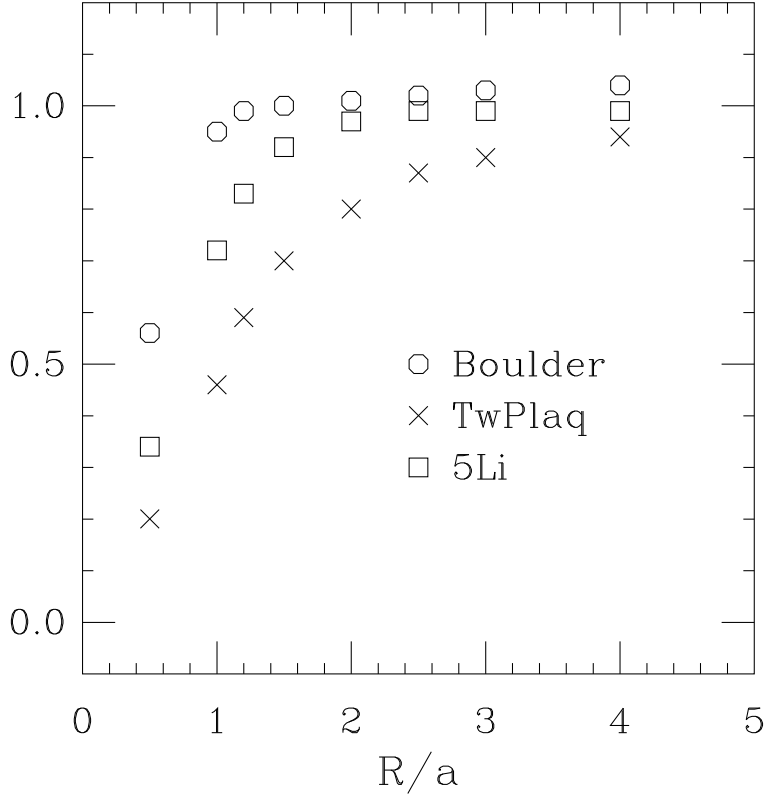}
\caption{Topological charge on artificial single instanton configurations as
a function of instanton radius, comparing three observables: the
traditional twisted plaquette operator, the Boulder operator
\protect\cite{ref:DHK.SU2} and the 5Li
operator\protect\cite{ref:fiveLi}.
\label{fig:canned_nosmooth}
}
\end{figure}

We first investigate how these operators perform with artificial
instantons.  We created a series of gauge configurations containing
a single instanton of varying radius and measured the topological charge with
each operator.  Results are plotted in Fig.~\ref{fig:canned_nosmooth}.
As expected \cite{ref:DHK.SU2} the twisted plaquette operator tends to
underestimate the topological charge for small instantons.  The 5Li
operator does better.  The optimization of the Boulder operator is
apparent.

\begin{figure}[ht]
 \epsfxsize=100mm
 \epsfbox{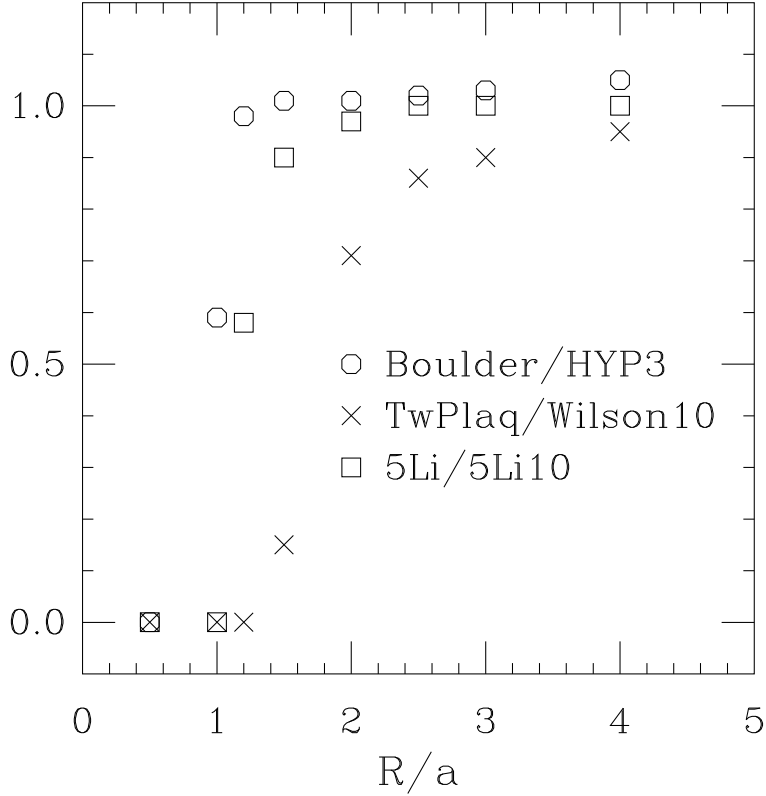}
\caption{Same as Fig.~\protect\ref{fig:canned_nosmooth}, but after smoothing.
\label{fig:canned_smooth}
}
\end{figure}

Traditional cooling methods include minimizing the Wilson action
\cite{ref:Wilsoncool} and minimizing improved actions, such as the 5Li
action \cite{ref:fiveLi}, that provide better scaling as a function of
instanton radius, so are less likely to erase small instantons.
Minimization is done through a series of standard heatbath updates
(cooling sweeps) at small gauge coupling.  Hypercubic smoothing
\cite{ref:HYPblock} was designed to be gentle and local so as to
produce a smooth configuration with minimal distortion of the topology
\cite{ref:HYPpreserve}.  The smoothing process involves a series of
APE blocking steps \cite{ref:APEblock}, constrained to lie entirely
inside the hypercubes connected to the link being smoothed.
We use the smoothing coefficients optimized in Ref.~\cite{ref:HN.SU3}.

To see how cooling or smoothing affects the artificial instantons, we
processed them using these methods.  For the twisted plaquette
operator we cooled with ten Wilson gauge action updates, for the 5Li
operator, ten 5Li updates.  For the Boulder operator we smoothed with
three hypercubic blocking sweeps. The number of smoothing steps in
each case was chosen for stability of $\chi$ under further smoothing,
as we shall discuss below.  Results are shown in
Fig.~\ref{fig:canned_smooth}.  It is evident that small instantons
preserve their topological charge best with the Boulder/HYP method.

\begin{figure}[ht]
 \vspace*{5mm}
 \epsfxsize=140mm
 \epsfbox{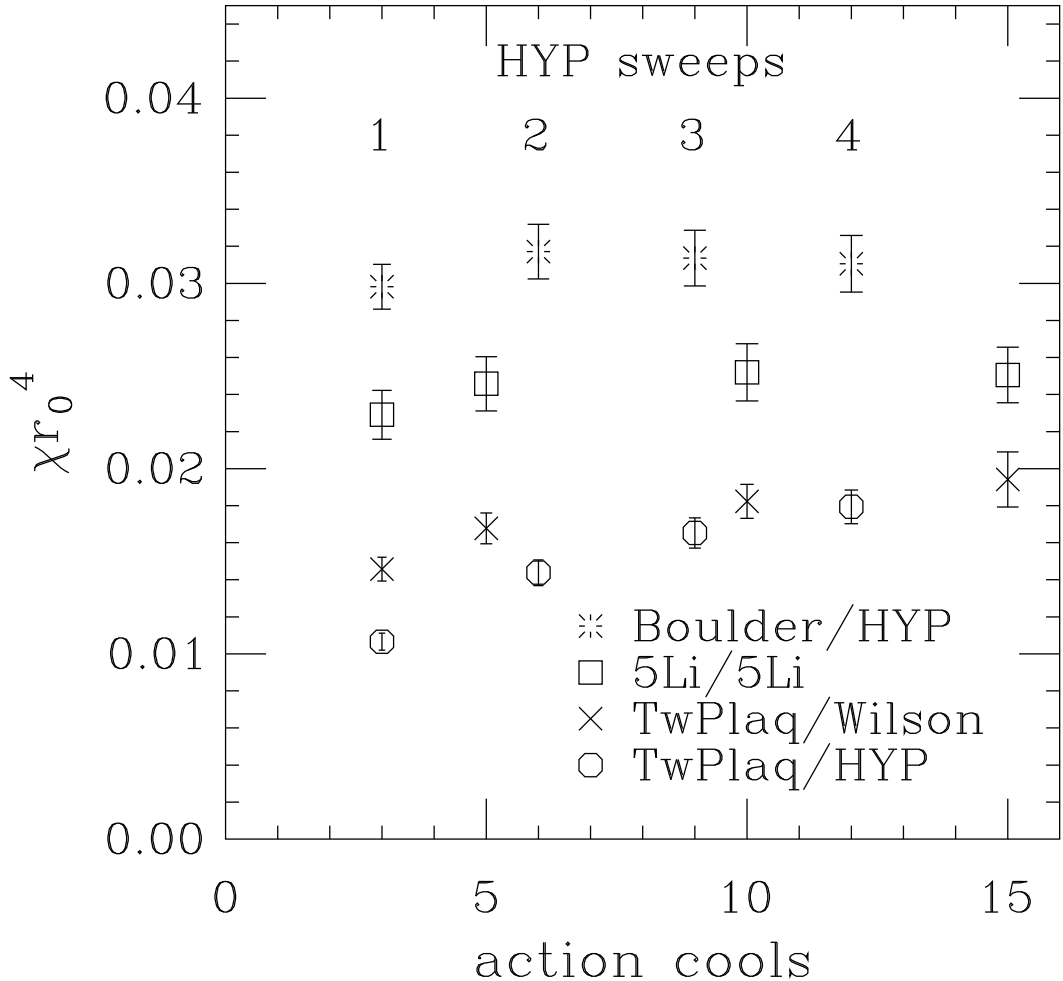}

\caption{Topological susceptibility as a function of smoothing or
cooling step on the $20^3 \times 64$ dataset with $am_{u,d} = 0.01$
and $am_s = 0.05$, comparing four techniques: the Boulder topological
charge operator with HYP smoothing \protect\cite{ref:HYPblock}, the
twisted plaquette operator \protect\cite{ref:twplq} with both Wilson
action cooling \protect\cite{ref:Wilsoncool} and HYP smoothing, and
the 5Li operator with 5Li cooling \protect\cite{ref:fiveLi}.  (Results
are expressed in units of $r_0$, the Sommer
parameter\protect\cite{ref:Sommer}).  The arbitrary scale conversion counts
three cooling steps for one HYP smoothing sweep.  Susceptibilities are
measured on subvolumes as explained in Sec.~\protect\ref{sec:results}.
\label{fig:smoothing_op} }
\end{figure}

Next we examine how the three methods perform on one of the gauge
ensembles in our study, namely the $20^3 \times 64$ $a = 0.12$ fm set
with quark masses $am_{u,d} = 0.01$ and $am_s = 0.05$
\cite{ref:dataset}.  We measured the topological susceptibility as a
function of cooling or smoothing step and compared the results in
Fig.~\ref{fig:smoothing_op}.  We see that it is reasonable to read off
the Boulder/HYP susceptibility after three HYP sweeps and the
TwPlaq/Wilson and 5Li/5Li susceptibilities after ten cooling sweeps.
We made these arbitrary choices in an effort to compromise between
preserving small instantons and reaching stability in the observable.
We have tested these choices in a few cases and find within
statistical errors that our results are insensitive to increasing
these values by a factor of two or three.  See also
Ref.~\cite{ref:HetrickLat02}.

It is clear that at this lattice spacing the TwPlaq/Wilson method
gives a lower susceptibility than the other methods. The 5Li/5Li
result is closer to but still lower than the Boulder/HYP result. To
determine to what extent the difference in TwPlaq/Wilson is
attributable to the observable and what to the cooling method, we also
measured the TwPlaq susceptibility on HYP smoothed lattices.  The
result (TwPlaq/HYP) shown in Fig.~\ref{fig:smoothing_op} is quite
close to the TwPlaq/Wilson result.  So the choice of operator appears
to account for the largest discrepancy.

We have also measured the topological susceptibility on the companion
quenched $20^3 \times 64$ $a = 0.12$ fm ensemble using three methods
and found similar inequalities: TwPlaq/Wilson gave $\chi r_0^4 =
0.036(2)$; 5Li/5Li, 0.051(3); and Boulder/HYP, 0.054(3).

\begin{figure}[ht]
 \epsfxsize=140mm
 \epsfbox{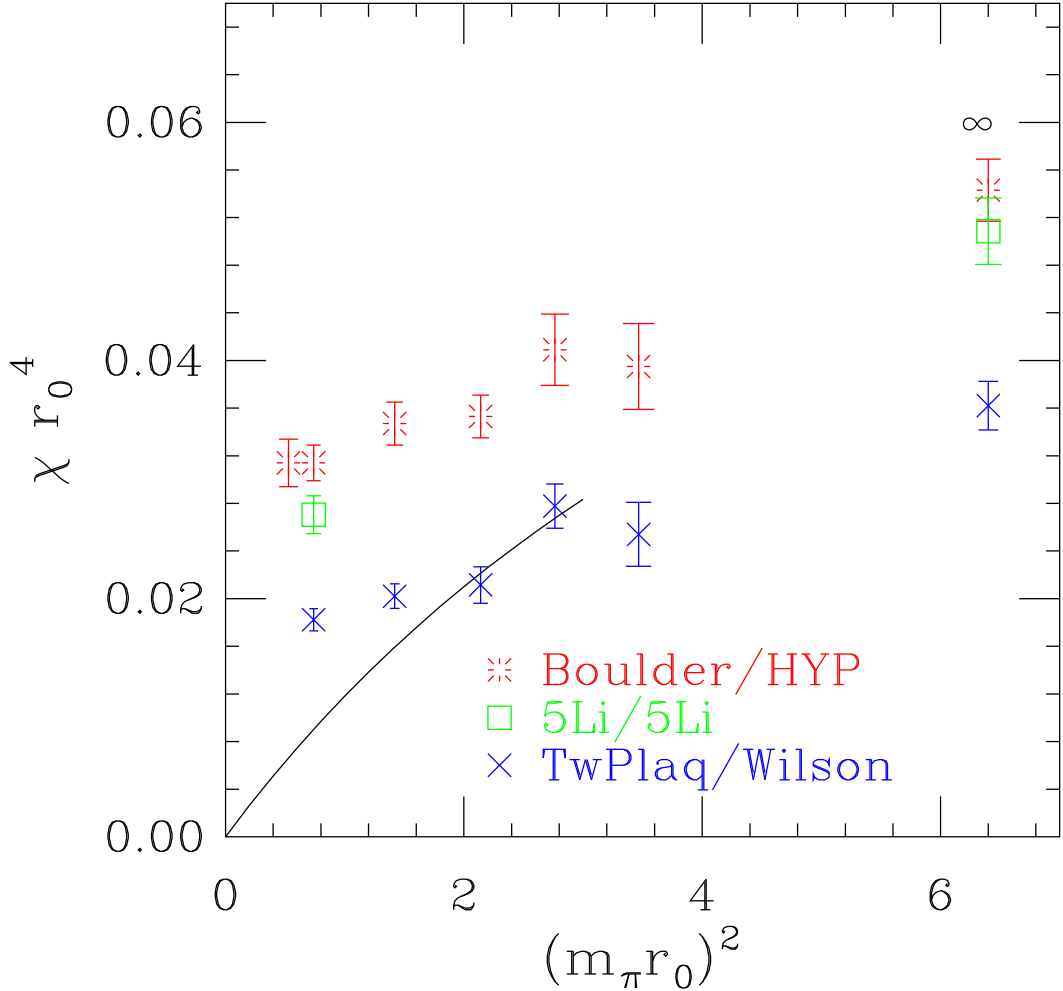}

\caption{Topological susceptibility {\it vs} pion mass squared in
units of $r_0$ on the $a = 0.12$ fm lattices, comparing the twisted
plaquette plus Wilson method method and the Boulder plus hypercubic
blocking method.  Also shown are two results for the 5Li/5Li method.
The solid line shows the prediction of leading order chiral
perturbation theory.  The quenched result is shown at the extreme
right.  The susceptibility is measured on subvolumes, as explained in
Sec.~\protect\ref{sec:results}.
\label{fig:chi_vs_mpi2_nf21_compare}
}

\end{figure}

Finally we considered the entire $20^3 \times 64$ dataset with $a
\simeq 0.12 \text{ fm}$ tabulated in
Table~\ref{tab:chi_vs_mpi2_nf21_asqtad}.  Taking ten cools and three
HYP sweeps for the comparison, we plot the result in
Fig.~\ref{fig:chi_vs_mpi2_nf21_compare}.  Throughout the entire mass
range the TwPlaq/Wilson susceptibility is about 2/3 the Boulder/HYP
value.

Since all three methods are expected to give the same continuum limit,
as they do for quenched QCD with the Wilson plaquette
action \cite{ref:TeperLat99}, the discrepancy we observe at lattice
spacing $a = 0.12$ fm must be due lattice artifacts. To decide what the
continuum limit is and which method is closer to it, one should do a
detailed scaling study.  We present results of a partial study in
Sec.~\ref{sec:results}, but here attempt to interpret the differences
based on our observations on smooth artificial instantons:

These discrepancies are of a magnitude that would be expected from our
measurements of the charge of artificial instantons of intermediate
size.  The average instanton radius is expected to be approximately
0.3 fm \cite{ref:ShuryakSchafer} or 2.3a on these $a = 0.12$ fm
lattices with significant contributions from radii as small as one
lattice unit.  From Fig.~\ref{fig:canned_smooth} where ideally $Q = 1$
we see that the TwPlaq/Wilson method underestimates the charge by
$Q(2.3a) = 0.85$ after cooling, so one would expect an underestimate
of about $0.85^2 = 0.7$ in the susceptibility for the average
instanton.  For smaller instantons the TwPlaq/Wilson method gives
$Q^2(1.5a) = 0.03$ after cooling, instead of one.  From
Fig.~\ref{fig:smoothing_op}, we find a ratio of 0.58(4) between the
TwPlaq/Wilson10 and Boulder/HYP3 susceptibilities at $(m_\pi r_0)^2 =
0.738$.  The ratio is approximately the same throughout the entire
mass range.  For the 5Li method we have $Q^2(2.3a) = 0.96$ and
$Q^2(1.5a) = 0.80$.  By comparison the ratio 5Li/5Li to Boulder/HYP3
is 0.87(7) at $(m_\pi r_0)^2 = 0.738$.  Consequently, one may wonder
whether the apparent agreement at $a = 0.12$ fm between the
TwPlaq/Wilson method and chiral perturbation theory at $2 \le (m_\pi
r_0)^2 \le 3$ is the result of compensating errors.

\section{Results}
\label{sec:results}

We measured the topological susceptibility using the the Boulder/HYP
method on two sets of gauge configurations generated with three flavors
of light Asqtad quarks of varying masses, one set with lattice spacing
approximately 0.12 fm throughout and the other, 0.09 fm
\cite{ref:dataset}.  The corresponding matched quenched configurations
are also included.  The data sample is tabulated in Table
\ref{tab:chi_vs_mpi2_nf21_asqtad}.

Besides measuring the susceptibility on the entire lattice volume, we
increased our statistics by measuring on smaller subvolumes.  The
probability distribution follows a Gaussian in $Q$ with width
proportional to the volume.  The width is decreased as the volume is
decreased, leading to the same relative error in the determination of
the susceptibility for the same sample size on the smaller volume.  To
the extent that the subvolume measurements are uncorrelated, the
sample is effectively increased by a factor equal to the number of
subdivisions, so the error should decrease by the square root of this
number.  This process cannot be continued indefinitely, since we
should eventually discover strong correlations among adjacent
subvolumes.

Measuring the susceptibility on subvolumes may even be indicated for
actions and updating algorithms that give persistent global charge,
but fluctuating local charge densities.

Accordingly, we divided the lattices along the time dimension into
three hypercubic subvolumes mostly separated by a small unused space.
For the $28^3\times 96$ lattices these subvolumes of size $28^4$ were
constructed from imaginary time ranges $[0,27]$, $[32,59]$, and
$[64,91]$, and for the $20^3\times 64$ lattices, of size $20^4$ from
time ranges $[0,19]$, $[22,41]$, and $[44,63]$.  Since the boundary
condition on the subvolume is not periodic, this practice splits
instantons.  However, the rms charge in the subvolumes (in the range $
4< |Q| < 7$) seems large enough to ensure that the boundary effects
are not significant.

We measured the correlations in the topological charge history between
charges in different subvolumes.  That is, if we denote by $Q_{uk}$
the charge measured in subvolume $u$ on gauge configuration $k$ in a
data sample with $N$ gauge field configurations, we define the
correlation coefficient to be
\begin{equation}
   c_{uv} = \frac{1}{N}\sum_k Q_{uk}Q_{vk}/(|Q_u||Q_v|)
\end{equation}
where $|Q_u| = \sqrt{\langle Q_u^2\rangle}$ is the rms charge on the
subvolume $u$.  The 33 correlation coefficients for our entire data
set are roughly Gaussian distributed about zero with a mean of
$-0.004$ and width of 0.09.  Thus we feel confident that we may treat
the subvolume measurements as statistically independent observations.

\begin{figure}[ht]
 \epsfxsize=140mm
 \epsfbox{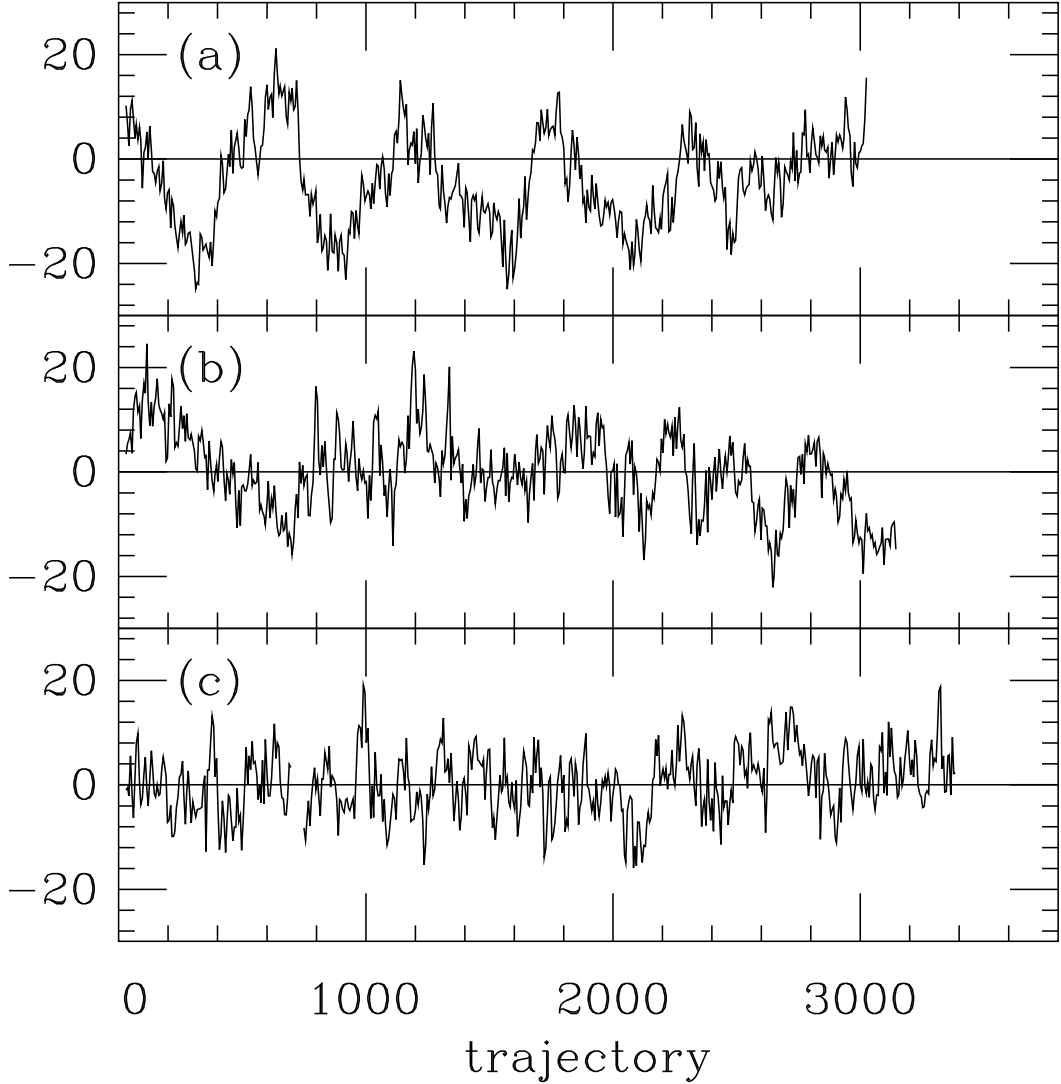}

\caption{Full volume topological charge after three HYP sweeps {\it
vs} molecular dynamics trajectory for the $28^3 \times 96$ data set
with (a) $am_{u,d} = am_s = 0.031$, (b) $am_{u,d} = 0.0124$, $am_s =
0.031$, and (c) $am_{u,d} = 0.0062$, $am_s = 0.031$.  In the last case
two separate time series are plotted with the second starting after
the break at 700 trajectories. \label{fig:q2896history} }
\end{figure}

\begin{figure}[ht]
 \epsfxsize=140mm
 \epsfbox{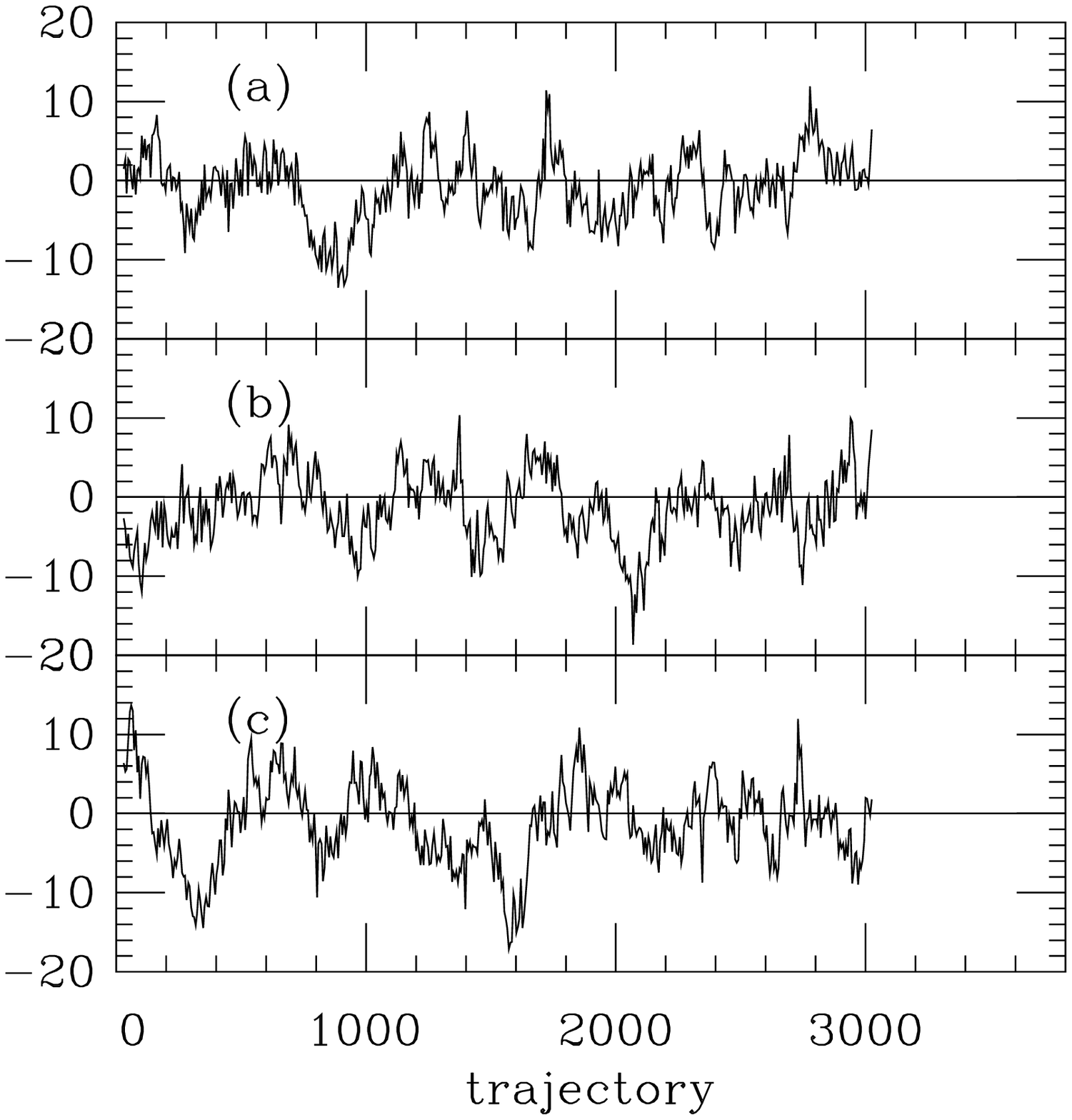}

\caption{Same as Fig.~\protect\ref{fig:q2896history}a, but measured on
three $28^4$ subvolumes defined by time slices (a) $[0,27]$, (b)
$[32,59]$ and (c) $[64,91]$.  \label{fig:qb2896m031} }
\end{figure}

The measurements are clearly correlated in Monte Carlo time.  The
Asqtad dynamical lattices were saved every sixth molecular dynamics
trajectory.  The $20^3 \times 64$ quenched lattices were saved every
tenth quasi-heatbath sweep and the $28^3 \times 96$ every fiftieth.
We made charge measurements on all available lattices.  A particularly
striking example is given by the time history for the total charge on
the $28^3\times 96$ lattice with three degenerate quark masses
$am_{u,d} = am_s = 0.31$, as shown in the upper panel of
Fig.~\ref{fig:q2896history} \cite{ref:suscept02}.  The horizontal
scale counts molecular dynamics trajectories.  Time histories for the
subvolume charges for the same dataset are shown in
Fig.~\ref{fig:qb2896m031} where the prominent oscillations are much
less evident.  Other time histories shown in the lower two panels of
Fig.~\ref{fig:q2896history} do not show such a striking effect. On the
$28^3 \times 96$ datasets the autocorrelation length, measured on the
subvolumes by summing the autocorrelation coefficient, ranges from
approximately 10 trajectories for $am_{u,d} = 0.0062$ to 35
trajectories  at $am_{u,d} = 0.031$.  Charge measurements on the
corresponding quenched lattices are only slightly correlated.  On the
coarser $20^3 \times 64$ lattices we find weaker autocorrelation, but
still roughly monotonically increasing with quark mass from fewer than
six trajectories at $am_{u,d} = 0.007$ to ten trajectories at
$am_{u,d} = 0.050$.  This trend appears to be contrary to some
expectations \cite{ref:Boyd}.

\begin{figure}[ht]
 \epsfxsize=140mm
 \epsfbox{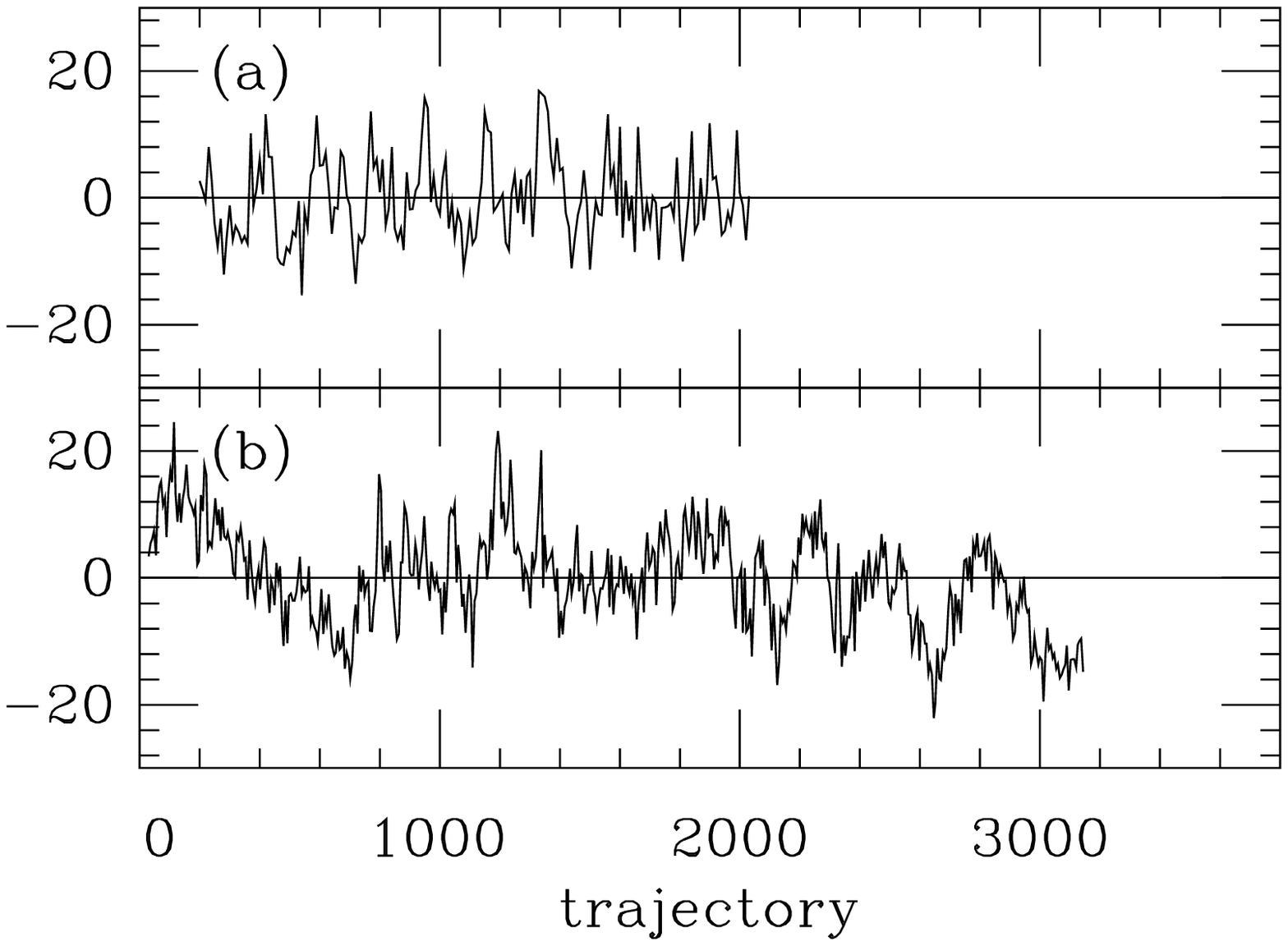}

\caption{(a) Topological charge after three HYP sweeps {\it vs} molecular
dynamics trajectory for the conventional unimproved staggered fermion
action on a $24^3 \times 64$ data set with two degenerate quark flavors
$am_{u,d} = 0.01$, compared with (b) the
result from the Asqtad action from Fig.~\protect\ref{fig:q2896history}(b).
\label{fig:thinvsfat}}
\end{figure}

The autocorrelations in topological charge found with the Asqtad
action and $2+1$ flavors of quarks appear to be longer than those
found with the conventional thin link staggered fermion action and two
flavors.  Shown in Fig.~\ref{fig:thinvsfat} is a comparison of the
topological charge history from the ensemble of
Fig.~\ref{fig:q2896history}b and an ensemble generated with the
conventional unimproved thin-link staggered fermion algorithm
\cite{ref:thin}.  These simulations were done at approximately the
same value of $(m_\pi r_0)^2$ (unimproved 1.06, improved 1.23) and
lattice spacing (unimproved 0.10 fm, improved 0.09 fm).  It is
apparent that the configurations decorrelate less rapidly with the
improved action and extra flavor.

Statistical errors in the topological susceptibility are determined by
taking the larger of the error corrected for autocorrelations and the
error obtained by extrapolating to infinite bin size.

\begin{figure}[ht]
 \epsfxsize=140mm
 \epsfbox{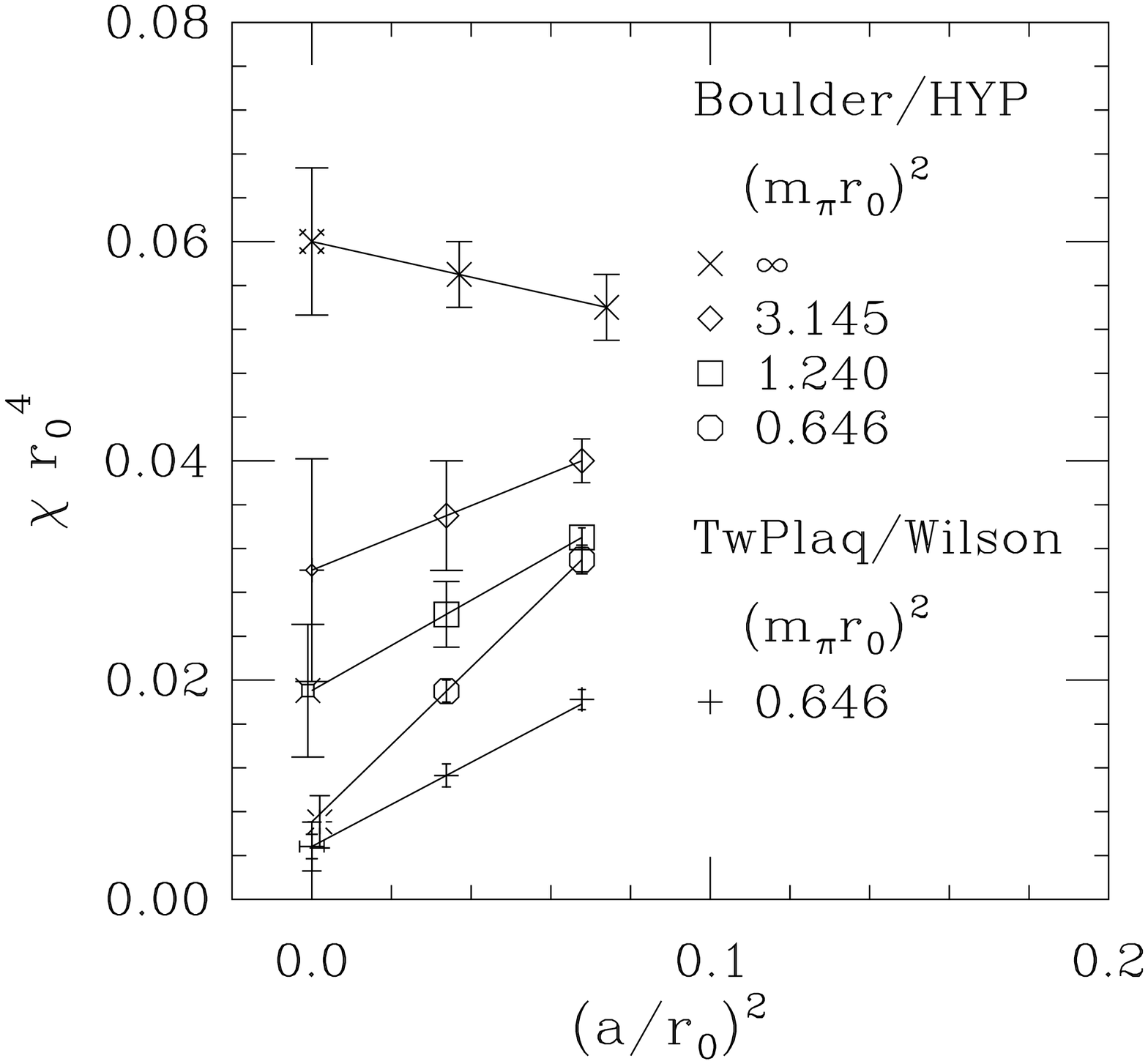}

\caption{Continuum extrapolation of the topological susceptibility
found with the Boulder/HYP and for the lightest quark mass the
TwPlaq/Wilson methods.  \label{fig:chi_vs_r0} }
\end{figure}

We have measurements of the topological susceptibility at two lattice
spacings.  Thus we may venture an extrapolation to the continuum
limit.  This is done by first interpolating in $(r_0 m_\pi)^2$ the
topological susceptibility on the coarse lattice to the three pion
mass values where we have measurements on the finer lattice.  We then
do a linear extrapolation at fixed $(m_\pi r_0)^2$ to zero $a^2/r_0^2$
as shown in Fig.~\ref{fig:chi_vs_r0}.  We see that the quenched
susceptibility rises with decreasing lattice spacing, but falls for
the unquenched lattices with a slope that increases with quark mass.

We have done a comparable extrapolation using the TwPlaq/Wilson
method, but with only one of the $a = 0.09$ fm dynamical ensembles.
Results are also shown in Fig.~\ref{fig:chi_vs_r0}.  Within the
limitations of these few points, we find satisfactory agreement
between the extrapolated values found with both methods.

\begin{figure}[ht]
 \epsfxsize=140mm
 \epsfbox{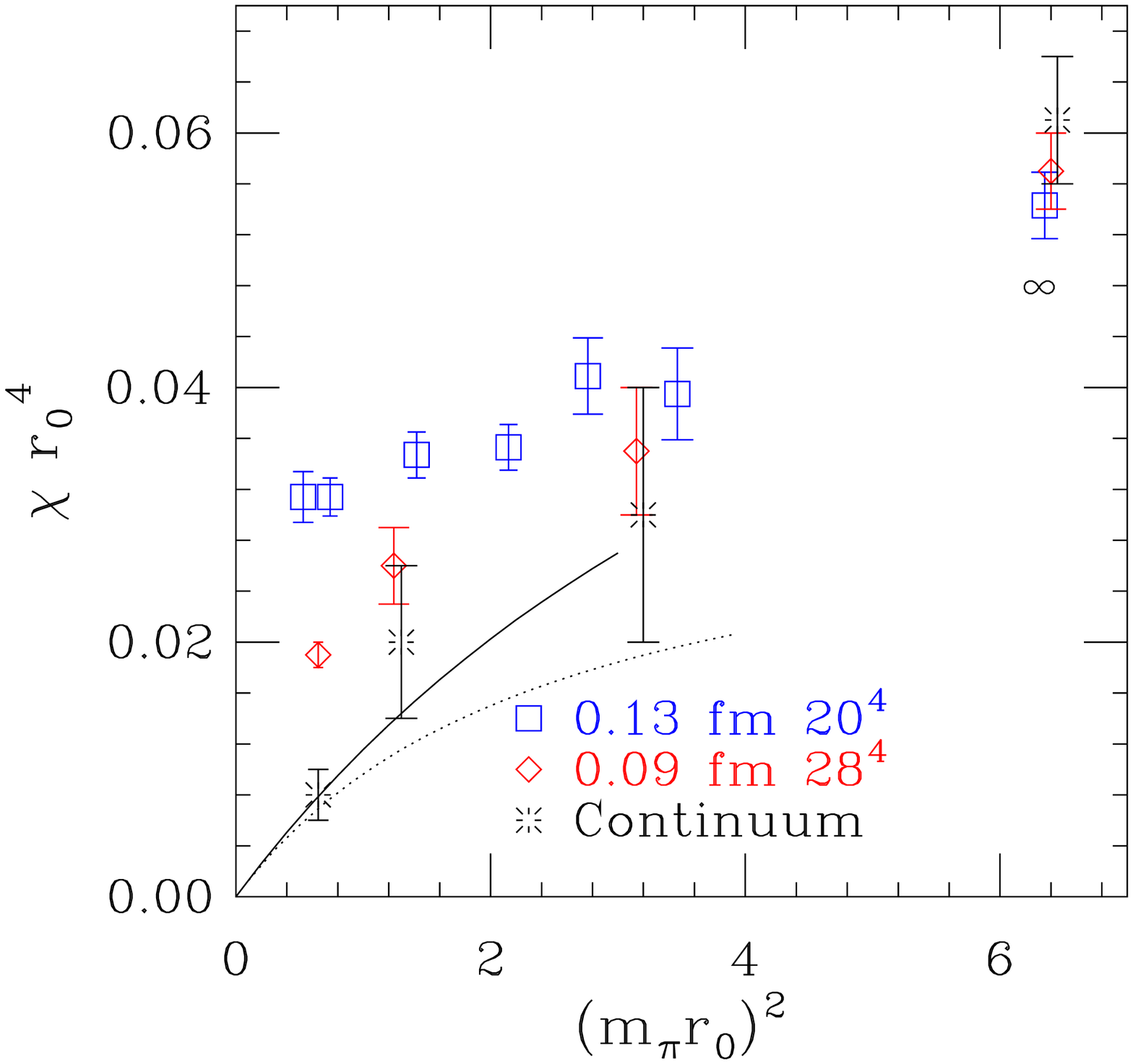}

\caption{Topological susceptibility {\it vs} pion mass squared in
units of $r_0$ with dynamical Asqtad quarks.  The solid line shows the
prediction of leading order chiral perturbation theory.  The dotted
line gives a phenomenological proposal for nonperturbative behavior
\protect\cite{ref:Durr2}.  The quenched result is shown at the extreme
right.  \label{fig:chi_vs_mpi2_nf21_asqtad} }

\end{figure}

Our main results for the topological susceptibility are summarized in
Table \ref{tab:chi_vs_mpi2_nf21_asqtad} and
Fig.~\ref{fig:chi_vs_mpi2_nf21_asqtad}.  We see that within errors the
continuum extrapolation gives reasonable agreement with the prediction
of leading order chiral perturbation theory, Eq (\ref{eq:chiral}).

\section{Discussion}
\label{sec:discussion}

The combined effect of the discretization artifacts discussed in
subsection \ref{subsec:artifacts} is that, depending on the method of
measurement, the lattice topological susceptibility $\hat{\chi}$ is
both additively and multiplicatively renormalized relative to the
continuum value, $\chi$ \cite{ref:twplq,ref:kronfeld}:
\begin{equation}
\hat{\chi}(a,m_q) = 
M(a,m_q)^2 \, \chi(m_q) + A(a,m_q),
\label{eq:renorm}
\end{equation}
where we understand the susceptibilities to be expressed in some
appropriate choice of physical units.  Having some indication of the
scaling behavior of these methods, we suggest a scenario for the
behavior of the functions $M$ and $A$ in Eq.~(\ref{eq:renorm}).

To suppress the ultraviolet fluctuations prior to measuring the
topological charge, it is necessary to smooth the gauge
configurations. In the quenched theory, smoothing eventually drives
$A$ to 0.  However, since fluctuations involving instantons of sizes
at or below the cutoff are always missed, one expects $M < 1$, at
least for operators that do not overestimate the charge of small
instantons.  Therefore, $\hat{\chi}^{\text{(qu)}} <
\chi^{\text{(qu)}}$.  Indeed, we see from
Fig.~\ref{fig:chi_vs_mpi2_nf21_compare} and the continuum
extrapolation in Fig.~\ref{fig:chi_vs_r0} that at $a = 0.12$ fm all
three methods underestimate the quenched susceptibility.

In full QCD the virtual fermions screen the topological charge, but
they screen everything they see, including dislocations.  Smoothing
removes dislocations and small instantons; those that are not involved
in the screening of more extended topological charges do not cause
trouble, but those that are leave behind the more extended (and now
less screened) charges that are relatively stable under smoothing.
This effect increases the susceptibility and shows up as an anomalous
$A$ that does not vanish with continued smoothing.

As we take the continuum limit in full QCD, we expect $M \to 1$, as in
the quenched case, and also $A \to 0$, as the lattice action, combined
with smoothing, gradually suppresses dislocations.  The two trends act
in opposite directions; to decide which dominates, we must consider
the quark mass dependence. There are two regimes. The first is for
large $m_q$, where $M$ dominates and $A$ is small. In this
``instanton-dominated'' region, the behavior is similar to the
quenched theory, and discretization effects lead to an underestimate
$\hat{\chi}(a,m_q) < \chi(m_q)$ at nonzero $a$. On the other hand, at
small quark mass the continuum susceptibility is small. The factor $M$
is expected to depend only weakly on $m_q$, so the lattice measurement
is dominated by $A$. In this ``dislocation-dominated'' regime there is
likely to be an enhancement $\hat{\chi}(a,m_q) > \chi(m_q)$ at nonzero
$a$.  Indeed, from Fig.~\ref{fig:chi_vs_mpi2_nf21_compare} and the
continuum extrapolations at $(m_\pi r_0)^2 = 0.646$ and infinity in
Fig.~\ref{fig:chi_vs_r0} we see that at $a = 0.12$ fm all three
methods overestimate the susceptibility expected at the nearby point
$(m_\pi r_0)^2 = 0.738$ and underestimate it at infinite quark mass.
With the Boulder/HYP method, our results suggest that the scaling
slope decreases as the quark mass is increased.

Given that the mean gauge action decreases by orders of magnitude
under smoothing at $a \simeq 0.1 \text{ fm}$, it is reasonable to
assume that dislocations are abundant and participate significantly in
screening more extended topological charges.  Erasing them then
contributes to $A$.  The number of instantons is not a strong function
of quark mass.  (The Wilson gauge action after twenty cooling sweeps
is dominated by instantons, so gives a measure of their number.  We
find that the mean action density on the coarse lattices varies by
less than 10\% across the dynamical and quenched ensembles in this
study).  We thus expect $A$ to vary at most weakly with the quark mass
in the dislocation-dominated region of quark mass.  Our results for
the $a = 0.12$ fm lattices plotted in
Fig.~\ref{fig:chi_vs_mpi2_nf21_compare} show that for all methods the
susceptibility levels off below $(m_\pi r_0)^2 \simeq 2$.

But leveling off can also be attributed to shortcomings of the lattice
fermion formulation itself.  The would-be zero modes of the staggered
lattice Dirac matrix are not at exactly zero. If the deviation is
comparable to the virtual quark mass, some of the topological modes
are not properly screened.  Consequently, $\hat{\chi}(a,m_q)$ does not
vanish as it should as the quark mass is reduced.  The chiral limit is
then governed by the dependence of $A$ on $m_q$.

\section{Conclusions}
\label{sec:conclusions}

We have measured the topological susceptibility on lattices generated
with Asqtad improved staggered fermions of varying mass and two
lattice spacings, $a = 0.12$ fm and $a = 0.09$ fm.  We have compared
three methods for measuring the topological charge on these lattices
and selected the Boulder method with hypercubic blocking, since it
appears best capable at these lattice spacings of preserving small
instantons.  We show in one comparison at $a = 0.12$ fm that the
largest difference between the Boulder/HYP and TwPlaq/Wilson methods
can be attributed to the operator, rather than the smoothing method.
We find that at both lattice spacings there is clear evidence that
dynamical quarks suppress topological fluctuations and the suppression
increases with decreasing quark mass and with decreasing lattice
spacing.  However, at fixed nonzero lattice spacing, lattice artifacts
at small quark masses are still substantial, and the susceptibility
does not decrease as expected from chiral perturbation theory at small
quark masses.  Nevertheless, an $O(a^2)$ extrapolation of our data to
zero $a$ gives results that are consistent with the leading order
prediction.  Within the limitations of our statistics, this is the
first study to show satisfactory agreement with the predictions of
chiral perturbation theory at quark masses much smaller than the
strange quark mass.

\section*{Acknowledgments}

We are grateful to Philippe de Forcrand for providing us his computer
code for the 5Li/5Li measurements.  Computations were performed at
LANL, NERSC, NCSA, ORNL, PSC, SDSC, FNAL, the CHPC (Utah), the Indiana
University SP and the Netra cluster (University of the Pacific).  This
work is supported by the U.S. National Science Foundation under grants
PHY01-39929 and PHY00-98395 and the U.S.~Department of Energy under
contracts DE-FG02-91ER-40628, DE-FG03-95ER-40894, DE-FG02-91ER-40661,
and DE-FG03-95ER-40906.  A.H\@. is supported by a Royal Society
University Research Fellowship.

\begin{table}
\begin{ruledtabular}
\caption{Topological susceptibility {\it vs} pion mass squared in
units of $r_0$.  The result $\chi_V$ is the susceptibility computed on
the full lattice volume.  The result $\chi$ is computed on three
subvolumes for improved statistics.} 
\label{tab:chi_vs_mpi2_nf21_asqtad}
\begin{tabular}{lllrlll}
 \multicolumn{7}{c}{$20^3\times 64$ ($a \approx .12$ fm)} \\
 $\beta$ & $am_{u,d}$ & $am_s$    & cfgs  & $(m_\pi r_0)^2$ & $\chi_V r_0^4$ & $\chi r_0^4$ \\
\hline                          
  6.76   & 0.007     & 0.050    & 446   & 0.529    & 0.031(2) & 0.0314(20) \\
  6.76   & 0.010     & 0.050    & 658   & 0.738    & 0.029(3) & 0.0314(15) \\
  6.79   & 0.020     & 0.050    & 486   & 1.420    & 0.033(2) & 0.0345(18) \\
  6.81   & 0.030     & 0.050    & 564   & 2.141    & 0.030(3) & 0.0353(18) \\
  6.83   & 0.040     & 0.050    & 351   & 2.764    & 0.042(4) & 0.0409(30) \\
  6.85   & 0.050     & 0.050    & 425   & 3.467    & 0.039(4) & 0.0395(36) \\
  8.00   & $\infty$  & $\infty$ & 409   & $\infty$ & 0.055(6) & 0.0543(28) \\
\hline
 \multicolumn{7}{c}{$28^3\times 96$ ($a \approx .09$ fm)} \\
 $\beta$ & $am_{u,d}$ & $am_s$    & cfgs \\
\hline
  7.09   & 0.0062    & 0.031    & 534   & 0.646    & 0.0155(16) & 0.0193(15) \\
  7.11   & 0.0124    & 0.031    & 520   & 1.240    & 0.023(5)   & 0.0260(27) \\
  7.18   & 0.031     & 0.031    & 500   & 3.145    & 0.043(9)   & 0.0351(51) \\
  8.40   & $\infty$  & $\infty$ & 416   & $\infty$ & 0.050(5)   & 0.0569(26) \\
\hline
\end{tabular} 
\end{ruledtabular}
\end{table}
\end{document}